\documentclass[aps,prb,reprint,superscriptaddress,floatfix]{revtex4-1}

\usepackage{amssymb,amsmath,bm,graphicx}
\usepackage{ifpdf,color,mathrsfs}

\graphicspath{{./fig/}}

\definecolor{darkblue}{rgb}{0,0,.6}
\definecolor{darkgreen}{rgb}{0,0.5,0}

\ifpdf
  \usepackage{epstopdf}
  \usepackage[pdftex,unicode,
  pdfstartview={FitH},pdfborder={0 0 0}]{hyperref}
  \usepackage{hypcap}
\else
  \usepackage[hypertex]{hyperref}
\fi
\hypersetup{
  bookmarksnumbered = true,
  colorlinks = true, linkcolor = darkblue,
  citecolor = darkblue, filecolor = darkblue,
  menucolor = darkblue, urlcolor = darkblue
}

\newcommand{\iu}{i} % {\mathrm{i}}  % imaginary unit i
\newcommand{\de}{d} % {\mathrm{d}}  % differential
\newcommand{\ee}{e} % {\mathrm{e}}  % Euler number

\let\Re\undefined

\DeclareMathOperator{\Re}{Re}

\newcommand{\ket}[1]{\ensuremath{\left| #1\right\rangle}}

\begin{document}

% \title{Highly nonlinear dynamics of resonantly excited charge carriers}
% \title{The role of intraband motion in coherent control of resonantly excited charge carriers}
% \title{Optical-field control of resonantly excited charge carriers}
% \title{The effect of laser acceleration on resonant excitation of charge carriers}
% \title{The effect of intraband dynamics on strong-field resonant excitation of charge carriers}
% \title{Strong-field coherent control in solids: the role of intraband dynamics}
\title{Strong-Field Resonant Dynamics in Semiconductors}

\author{Michael S.~Wismer}
\email[]{michael.wismer@mpq.mpg.de}
\affiliation{Max-Planck-Institut f\"ur Quantenoptik, Hans-Kopfermann-Str. 1, 85748 Garching, Germany}

\author{Stanislav Yu.~Kruchinin}
\email[]{stanislav.kruchinin@mpq.mpg.de}
\affiliation{Max-Planck-Institut f\"ur Quantenoptik, Hans-Kopfermann-Str. 1, 85748 Garching, Germany}

\author{Marcelo Ciappina}
\affiliation{Max-Planck-Institut f\"ur Quantenoptik, Hans-Kopfermann-Str. 1, 85748 Garching, Germany}

\author{Mark I.~Stockman}
\email[]{mstockman@gsu.edu}
\affiliation{Center for Nano-Optics (CeNO) and Department of Physics and Astronomy, Georgia State University, Atlanta, Georgia 30340, USA}

\author{Vladislav S.~Yakovlev}
\email[]{vyakovlev@gsu.edu}
\affiliation{Max-Planck-Institut f\"ur Quantenoptik, Hans-Kopfermann-Str. 1, 85748 Garching, Germany}
\affiliation{Center for Nano-Optics (CeNO) and Department of Physics and Astronomy, Georgia State University, Atlanta, Georgia 30340, USA}

\date{\today}

\begin{abstract}
We predict that a direct bandgap semiconductor (GaAs) resonantly excited by a strong ultrashort laser pulse exhibits a novel regime: kicked anharmonic Rabi oscillations (KARO).
In this regime, Rabi oscillations are strongly coupled to intraband motion, and interband transitions mainly take place during short times when electrons pass near the Brillouin zone center where electron populations undergo very rapid changes.
Asymmetry of the residual population distribution induces an electric current controlled by the carrier-envelope phase.
The predicted effects are experimentally observable using photoemission and terahertz  spectroscopies.
\end{abstract}

\pacs{72.20.Ht,78.47.-p}
%\keywords{}

\maketitle

\section{Introduction}
\begin{figure}[!tb]
  \centering
  \includegraphics[width=\columnwidth,height=0.7\textheight,keepaspectratio]{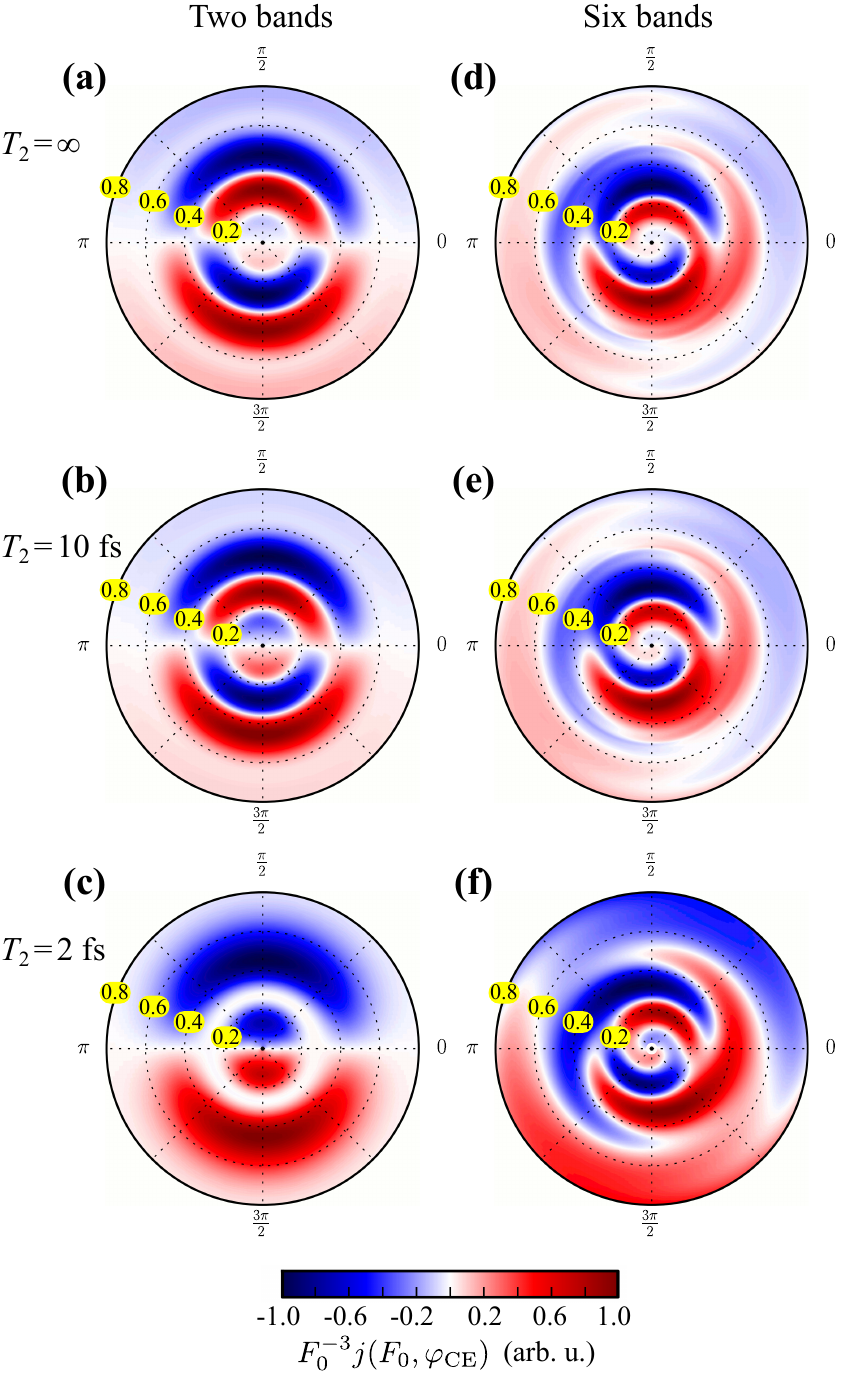}
  \caption{(Color online) The residual current density $j(F_0, \varphi_{\text{CE}})$.
    In these diagrams, the distance to the origin corresponds to the pulse amplitude $F_0$, which varies from zero to 0.8~V/{\AA}, while the angle to the horizontal axis encodes the carrier-envelope phase $\varphi_{\text{CE}}$.
    The color coding of $F_0^{-3} j(F_0, \varphi_{\text{CE}})$ is individually normalized for each diagram.
%    The solid green curve represents $\varphi_{\text{CE}}$ that maximizes the residual current.
    Panels (a)--(c) show two-band results (1 VB, 1 CB), while panels (d)--(f) display six-band  (3 VBs, 3 CBs) calculation results.
    Each horizontal pair of plots corresponds to a certain value of dephasing time $T_2$ as indicated by the labels.}
\label{Figure1}
\end{figure}

Effect of strong electric fields on crystalline solids was long considered important as introduced by Zener~\cite{Zener_1934_PRS_145_523} and developed by Keldysh~\cite{Keldysh_1965_JETP_20_1307}.
Recently, a novel field studying interaction of few-cycle high-intensity (with fields reaching or exceeding internal fields in the matter) optical pulses with solids has attracted a great deal of attention~\cite{Ghimire2011, Schiffrin2013, Schultze2013, Krausz2014, Schultze2014, Schubert2014, Luu2015, Vampa2015, Hohenleutner2015, Corkum_2015_PRL_115_193603}.
One of the most significant directions of the recent research has been strong-field interaction of transparent solids with ultrashort pulses.
In that case, the carrier frequency $\omega_0$ and the pulse bandwidth $\Delta\omega$ are well within the solid's bandgap: $\hbar \omega_0, \hbar\Delta\omega\ll E_g$.
For ultrashort pulses, light-induced processes in transparent solids develop and evolve necessarily within the optical cycle~\cite{Schiffrin2013, Schultze2013, Schultze2014}; they are largely reversible, and the characteristic response time $\tau$ is determined by the band gap: $\tau\gtrsim \hbar/E_g$~\cite{Krausz2014}.
Consequently, these subcycle processes depend on the carrier-envelope phase (CEP)~\cite{Jones2000}.

The CEP controllability of electron dynamics in the case of resonant excitation, $\hbar \omega_0 \sim E_g$, has so far mostly been studied in the context of carrier-wave Rabi flopping (CWRF)~\cite{Hughes1998, Muecke2001, Muecke2002, Muecke2004}, where the motion of charge carriers within bands was neglected, as well as in the context of perturbative $\omega+2\omega$ interference~\cite{Atanasov1996, Fortier2004}, where a relatively weak field limits the scope of the intraband motion.

Here we predict that intraband electron motion in strong resonant fields determines electron excitation of a semiconductor solid.
These strong-field resonant electron dynamics cause a residual ballistic electric current that is controllable by the CEP.
This current has a nontrivial dependence on the field amplitude, exhibiting deep oscillations and direction reversals.

\section{Model}

We solve the length-gauge optical Bloch equations with intraband displacement terms~\cite{Golde2008, Schubert2014}:
\begin{multline}
  \label{eq:model}
  \frac{\partial}{\partial t} \rho_{i j} =
  \left[
    \frac{\delta_{i j} - 1}{T_2} + \frac{\iu}{\hbar} (E_i - E_j)
  \right]
  \rho_{i j} \\+
  \frac{1}{\hbar} \bm{F}(t)
  \bigl(
    e\nabla_{\bm{k}} \rho_{i j} - \iu [\hat{\bm d}, \hat\rho]_{i j}
  \bigr).
\end{multline}
Here, $\rho(\bm{k}, t)$ is a density matrix, its diagonal elements, $n_i(\bm{k}, t) = \rho_{i i}(\bm{k}, t)$, are probabilities to find an electron with crystal momentum $\bm{k}$ in band $i$, $T_2$ is dephasing time introduced phenomenologically; $e > 0$ is the elementary charge, and $\bm{d}_{i j}(\bm{k}) = e\langle \psi_i(\bm{k})| \iu \nabla_{\bm{k}} | \psi_j(\bm{k}) \rangle$ are dipole matrix elements that form matrix $\hat{\bm d}$.
We assume that the electric field of the laser pulse \emph{in the medium} $\bm{F}(t)$ is linearly polarized along the $\Gamma-\text{X}$ direction in the Brillouin zone of GaAs, where the X point is at $k_{\text{max}} = 1.11\,\text{\AA}^{-1}$.
This choice eliminates second-order nonlinear effects, in particular, optical rectification~\cite{Zhang1992}.
We will denote the field projection on the $\Gamma-\text{X}$ direction as $F(t)$ and its amplitude as $F_0$.
The specific form of $F(t)$ is described in Appendix B.
This is a 5-fs pulse with a central (carrier) frequency of $\hbar\omega_0 = E_g = 1.55$~eV.

We obtained the band energies $E_i(\bm{k})$ and the transition matrix elements $\bm{d}_{i j}(\bm{k})$ using the Wien2K code~\cite{Wien2k}.
We found that using three conduction bands (CBs) and three valence bands (VBs) was sufficient for convergence. % with a good accuracy.

\section{Results}

Let us make a few relevant estimates.
Nondestructive measurements on GaAs with few-cycle pulses were reported for $F_0=0.4$~V/{\AA} (a peak intensity of $2 \times 10^{12}$~W/cm$^2$), where the onset of CWRF was observed~\cite{Muecke2001,Muecke2004}.
This field is much smaller than that required to accelerate an electron from the $\Gamma$ point ($\bm{k} = 0$) to the boundary of the Brillouin zone, which is $F_0 = 0.9$~V/{\AA} for $\lambda_0 = 800$~nm.
Using $d= 5\ e\cdot \text{\AA}$~\cite{Muecke2001}, we estimate the ratio of the Rabi frequency $\Omega_R = d F_0/\hbar$ to the laser frequency as $\Omega_R / \omega_0 \approx F_0 / (0.3~\text{V/\AA})$.

\begin{figure}[t]
  \centering
  \includegraphics[width=0.95\columnwidth]{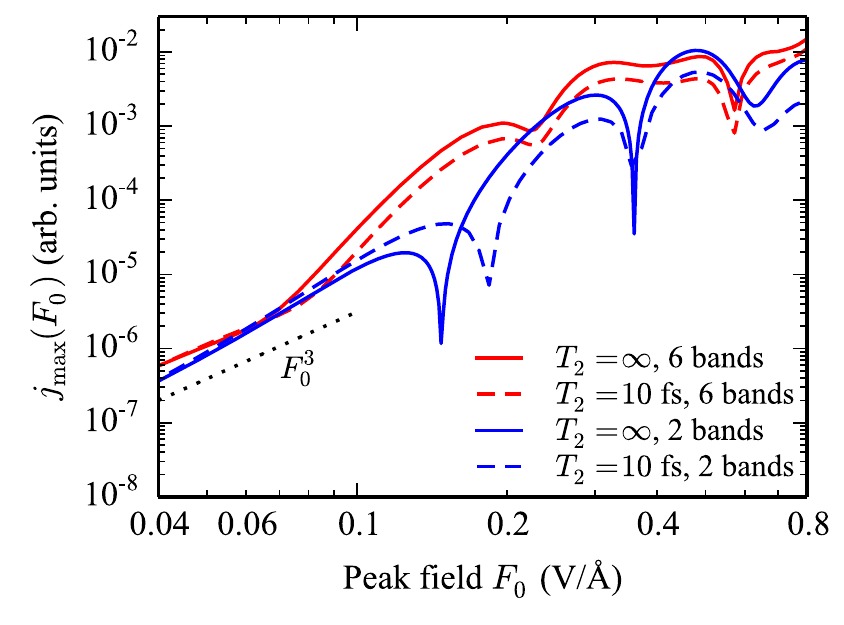}
  \caption{(Color online) The maximal value of the residual current density $j_{\text{max}}(F_0) = \max_{\varphi_{\text{CE}}}[j(F_0, \varphi_{\text{CE}})]$.
    The solid and dashed lines were obtained with $T_2 = \infty$ and $T_2 = 10$~fs, respectively.
    Red curves represent six-band calculations (3 VBs, 3 CBs), whereas blue curves show the two-band results (1 VB, 1 CB).}
\label{fig:current}
\end{figure}

Figure \ref{Figure1} shows the residual current density,
\begin{equation}
  \label{eq:intraband_current}
  j\left(F_0, \varphi_{\text{CE}}\right) =
  -\frac{2 e}{(2 \pi)^3}\sum_i \int_{\text{BZ}} \de^3 k\, n_i(\bm{k}, t_{\text{max}})
  \hat{\bm{e}}\bm{v}_i(\bm{k}),
\end{equation}
for the cases of two [Figs.~\ref{Figure1}(a)-(c)] and six [Figs.~\ref{Figure1}(d)-(f)] bands, as well as for different values of $T_2$.
When the field is weak, the photocurrent is excited due to the $\omega+2\omega$ interference~\cite{Atanasov1996,Fortier2004}.
In this case, it is known that $j_{\text{max}}(F_0) \propto F_0^3$---cf.\ Fig.~\ref{fig:current}.
This is due to the fact that the probability amplitudes of one- and two-photon processes are proportional to $F_0$ and $F_0^2$, respectively, while their interference makes a contribution proportional to $F_0^3$.
In Fig.~\ref{fig:current}, this cubic dependence breaks down for $F_0 \gtrsim 0.1$~V/{\AA}, which we visualize in Fig.~\ref{Figure1} by representing $F_0^{-3} j\left(F_0, \varphi_{\text{CE}}\right)$ with color coding.
In Fig.~\ref{Figure1}, the results obtained for two and six bands differ significantly, which is consistent with recent findings~\cite{Hohenleutner2015}.
However, they also share a few remarkable features.

First, we observe CEP-controlled light-induced residual current, which implies that it is due to ultrafast, subcycle processes.
The cases of no polarization relaxation (or, zero dephasing, $T_2=\infty$) [panels (a) and (d)] and fast dephasing [$T_2=10$~fs, panels (b) and (e)]  differ very little, which suggests that there is fast effective dephasing within the purely Hamiltonian system described by the Schr\"odinger equation.
Note that the fastest electron dephasing time in semiconductors (GaAs) was measured to be $T_2\sim 14$~fs \cite{Becker_PRL_1988}, which was consistent with theory \cite{Vu2004}. 
At the same time, recent experiments on high-harmonic generation in solids~\cite{Schubert2014, Luu2015, Vampa2015, Hohenleutner2015, Corkum_2015_PRL_115_193603} suggest that dephasing times in the strong-field regime may be on the order of femtoseconds, so we also present results for $T_2 = 2$~fs.
We note that $T_2$ has a stronger impact on the two-band results.

Second, for any chosen CEP, $j\left(F_0, \varphi_{\text{CE}}\right)$ changes its sign at certain values of $F_0$.
In the two-band model, the maximum magnitude of the current at any field amplitude is always obtained for the antisymmertic pulse ($\varphi_{\text{CE}} = \pm\pi/2$).
In contrast, for more realistic six-band calculations, the maximum current non-trivially depends on the CEP, which causes the appearance of ``vortices'' in panels (d)--(f).

Third, starting from $F_0 \sim 0.2$~V/{\AA}, the residual current is much stronger than that obtained by extrapolating the weak-field current according to the $\propto F_0^3$ law.
This fact is more clearly seen in Fig.~\ref{fig:current}, from which we also conclude that dephasing tends to reduce the magnitude of the residual current.

\begin{figure*}[!htb]
  \centering
  \includegraphics[width=0.9\textwidth,height=0.7\textheight,keepaspectratio]{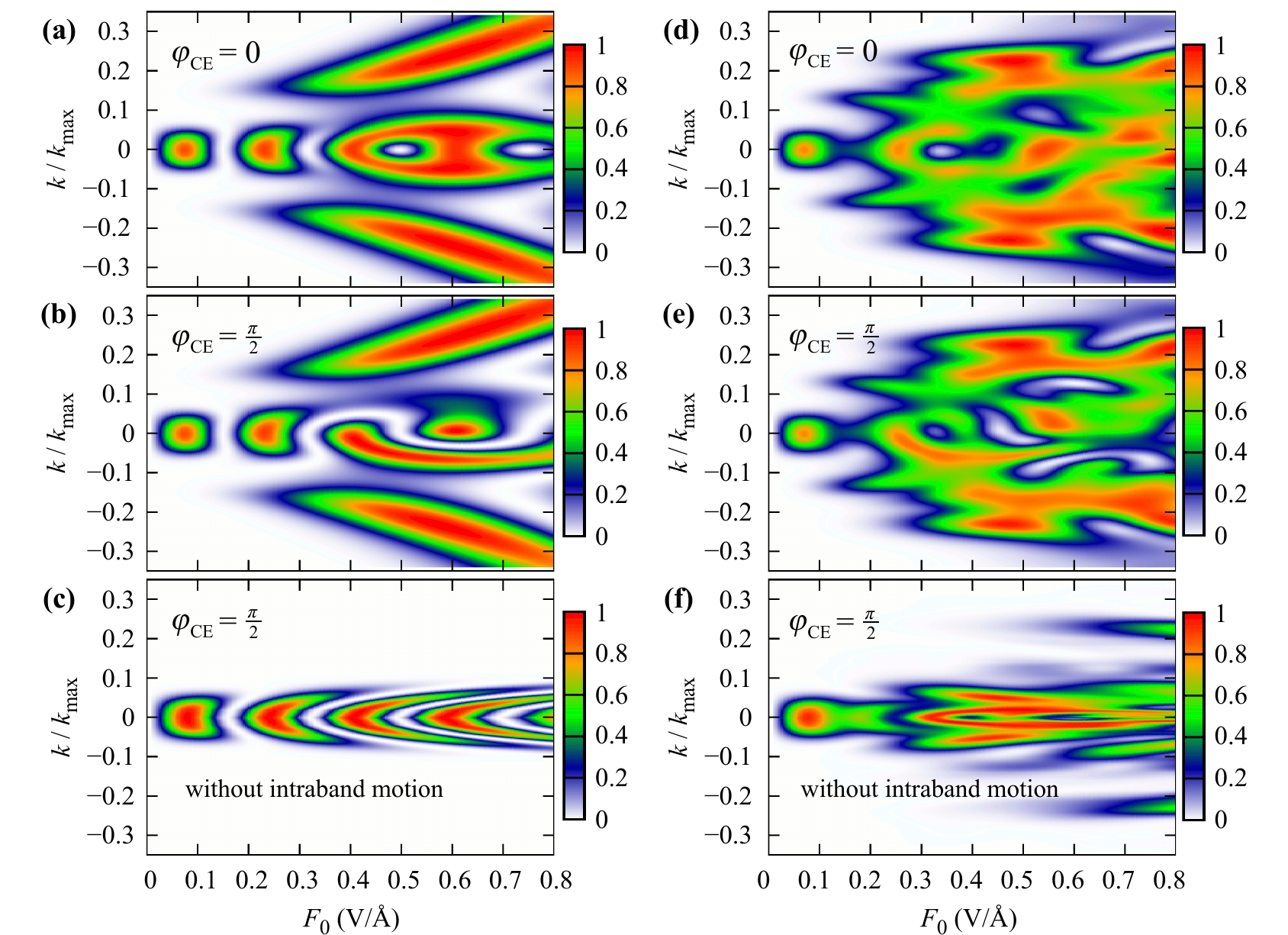}
  \caption{(Color online) The residual population of the lowest conduction band $n_{\text{c}_1}(k, t_{\text{max}})$ in simulations with two (a)--(c) and six (d)--(f) bands without dephasing ($T_2 = \infty$).
  The CEP of the laser pulse is $\varphi_{\text{CE}} = 0$ in panels (a), (d) and $\pi/2$ in the other plots.
  Panels (c), (f) display population distributions obtained without intraband motion.}
  \label{Figure3}
\end{figure*}
To gain more insight, we analyze in Fig.~\ref{Figure3} the residual population $n_{\text{c}_1}(k, t_{\text{max}})$ of the lowest conduction band (c$_1$).
Higher bands contribute significantly less to the electric current.
As we have already pointed out,  dephasing with a typical rate of $T_2 \gtrsim 10$~fs has a minor effect; thus we set $T_2= \infty$.
We start with the case of two bands displayed in Figs.~\ref{Figure3}(a)--(c) where it is easier to disentangle different processes.
For $\varphi_{\text{CE}} = 0$ [panel (a)], the $k$-resolved population is symmetric, so that the residual intraband current is close to zero, in accord with Fig.~\ref{Figure1}.
For $\varphi_{\text{CE}} = \pi/2$ [panel (b)], the residual conduction-band population becomes highly asymmetric in $\pm k$  in the region $F_0 \gtrsim 0.3$~V/{\AA} and $|k|/k_{\text{max}} \lesssim 0.1$.
As Fig.~\ref{Figure3}(c) demonstrates, intraband motion plays a key role here.
Eliminating it by dropping the term $\propto \bm F(t)\nabla_{\bm{k}} \rho_{i j}$ in Eq.~\eqref{eq:model} leads to a fully symmetric distribution of excitation probabilities and also eliminates the excited populations at $k\gtrsim 0.2 k_{\text{max}}$.

% In Fig.~\ref{Figure3}(a)--(c), we see high-contrast fringes where the population changes between 1 and 0.
% These fringes are due to quantum-mechanical interference of events of the carrier population that originate at different moments in time.
Interband transitions predominantly occur in the part of the Brillouin zone where the corresponding transition matrix elements $\bm{d}_{i j}$ have a large magnitude.
For transitions from valence bands to the lowest conduction band of GaAs, $|\bm{d}_{i j}(\bm{k})|$ are maximal at the $\Gamma$ point and sharply decrease for $k\gtrsim \Delta k = 0.1 k_0$ (see Appendix A).
According to the Bloch acceleration theorem~\cite{Bloch_1929_ZPhys_52_555}, the electron momentum in a given energy band, $\bm{K}(t)$, changes in time as $\bm{K}(t)=\bm{k}- e \hbar^{-1} \int_{t_0}^t\bm{F}(t^\prime) \de t^\prime$, where $\bm{k} = \bm{K}(t_0)$ is the initial momentum.
At a characteristic field of $ F\sim 0.4~\text{V/\AA}$, an electron at the $\Gamma$ point passes through the momentum range of interband transitions during time $\tau \sim \hbar \Delta k / (e F_0) \sim 0.2$~fs.
This time interval is much shorter than a half-cycle of both optical oscillations and Rabi oscillations.

\begin{figure}[!htb]
  \centering
  \includegraphics[width=\columnwidth]{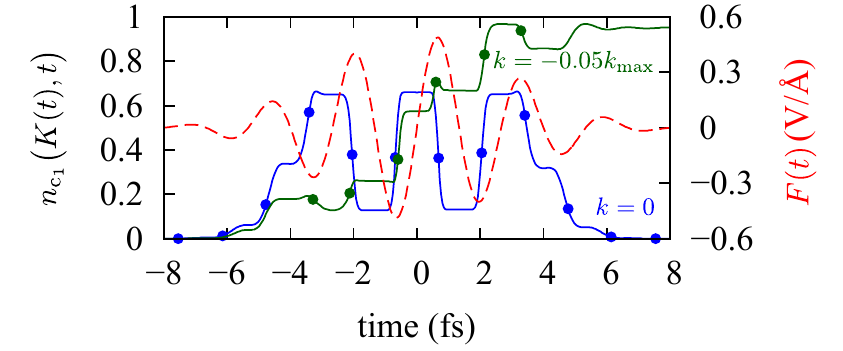}
  \caption{(Color online) 
    Time dependence of the CB population in the two-band simulation (blue and green curves).
    It is calculated along reciprocal-space pathways $K(t)$ that satisfy the acceleration theorem.
    The starting point is the $\Gamma$ point ($k = 0$) for the blue curve and $k=-0.05k_{\text{max}}$ for the green curve.
    The bold dots on the curves denote moments of passage in the vicinity of the $\Gamma$ point.
    The dashed red curve shows the electric field of the pulse ($F_0 = 0.5$~V/{\AA}, $\varphi_{\text{CE}} = \pi/2$).
}
\label{fig:Figure4}
\end{figure}

Thus we predict a novel regime for solids in strong resonant optical fields where the Rabi oscillations are excited by short and strong kicks during times when electrons pass the narrow momentum region $\sim \Delta k$ near the $\Gamma$ point.
These kicks are repeated twice per optical cycle causing strongly anharmonic (non-sinusoidal in time) Rabi oscillations.
We call this regime ``\emph{kicked anharmonic Rabi oscillations}'' (KARO).
It is illustrated in Fig.~\ref{fig:Figure4} where we show the time dependence of the lowest CB population for selected reciprocal-space pathways: $n_\mathrm{c_1}\bigl(\bm{K}(t),t\bigr)$ undergoes very rapid changes at the moments of the kicks, and it is nearly constant between them.
Such transitions can excite or de-excite electrons, and their overall outcome depends on the field amplitude and initial momentum $\bm{k}$, in turn defining whether there is a ``bright'' (high $n_\mathrm{c_1}$) or ``dark'' (low $n_\mathrm{c_1}$) fringe.
At the $\Gamma$ point ($k = 0$) in Fig.~\ref{fig:Figure4}, the blue curve clearly indicates alternation of such excitation/de-excitation events.
For $k \ne 0$ (the green curve), an electron passes near the $\Gamma$ point at non-equidistant moments of time, which affects the final population and current.

If the intraband motion is artificially switched off, the fringes in Fig.~\ref{Figure3}(c) repeat periodically, as expected for Rabi oscillations, and the distribution is precisely symmetric: $n_\mathrm{c_1}(k, t_{\text{max}})= n_\mathrm{c_1}(-k, t_{\text{max}})$.
In contrast, the KARO regime in Figs.~\ref{Figure3}(a) and (b) manifests itself by the fringe spacing increasing with $F_0$.
However, the most important signature of the KARO is evident asymmetry of the CB density distribution with respect to $k < 0$ vs.\ $k > 0$, which appears due to the intraband electron motion induced by an ultrashort pulse.
The number of Rabi cycles increases with the field amplitude; when the pulse duration corresponds to an integer number of full oscillations, the residual current switches its direction---cf.\ Fig.~\ref{Figure1}.
We interpret this switching of current as resulting from interference of electron pathways in reciprocal space.

The change of band populations between times $t_i$ and $t_f$ is determined by the field work, $W=\int_{t_i}^{t_f}\bm{F}(t) \bm{P}'(t)\,dt$, where $\bm{P}'(t)$ is the time derivative of the macroscopic polarization induced by both bound and free charges \cite{Landau_1984_8}.
In the KARO process, the field changes little during a single kick.
Consequently, a kick at time $t$ does work $\Delta W \approx \bm{F}(t) \Delta \bm{P}(t)$, where $\Delta \bm{P}$ is the corresponding polarization change.
Two kicks that promote electrons from band $j$ to band $i$ at times $t_1$ and $t_2$ interfere with each other according to the phase accumulated by the interband polarization between the kicks.
For a particular electron with initial crystal momentum $\bm{k}$, this phase is approximately given by (see Appendix D)
\begin{equation}
\label{eq:phase_shift}
  \Delta\phi_{ij}(\bm{k}) = \frac{1}{\hbar} \int_{t_1}^{t_2} \de t\, \Delta E_{ij} \bigl(\bm{K}(t)\bigr).
\end{equation}
It is analogous to the Volkov phase \cite{Volkov1935} in atomic physics.
Let us consider two pathways separated by an optical cycle: $t_2 - t_1 = 2\pi/\omega_0$.
When $\Delta\phi_{ij}(\bm{k}) = 2 \pi q$, where integer $q$ is the order of a nonlinear resonance, interference is constructive resulting in net excitation of electrons.

In Figs.~\ref{Figure3}(a)--(b), it is evident that a region of strongly nonlinear behavior occurs at $F_0\gtrsim 0.5$~V/\AA.
A nonlinear resonance of the lowest order ($q=2$) for electron wave packets excited near the $\Gamma$ point occurs for $\Delta\phi_{i j}(0) = 4 \pi$.
In a model with two parabolic bands, $\Delta E_{i j}(\bm{k}) \approx E_g + \hbar^2 k^2 / (2 \mu)$, where $\mu$ is reduced effective mass.
From Eq.~(\ref{eq:phase_shift}), we obtain the second-order resonance condition as $E_g + U_p = 2 \hbar \omega_0$, where $U_p = e^2 F_0^2 / (4 \mu \omega_0^2)$ is known as ponderomotive energy.
For GaAs, using $\mu = 0.05 m_0$, we obtain $F_0 = 0.3$~V/{\AA} for transitions from the light-hole VB to the lowest CB.
A more careful calculation taking band non-parabolicity into account yields $F_0 = 0.5$~V/{\AA}.

While we have concentrated above on a two-band model, real crystals contain a number of bands that can contribute to nonlinear optical phenomena in strong fields.
Figures~\ref{Figure3}(d)--(f) show that increasing the number of bands to six has a dramatic effect on the simulations for $F_0 \gtrsim 0.1$~V/{\AA}: reciprocal-space populations are perturbed in a highly irregular, quasi-stochastic way; they become significantly asymmetric at lower laser fields, and signatures of Rabi cycles at $\bm{k}=0$ are not as clearly visible even if intraband motion is neglected.
%Interestingly enough, the strong laser field is capable of coupling bands separated by energies far beyond the laser bandwidth.
% The presence of almost zero-population regions in Fig.~\ref{Figure3}(d)--(e) implies that in some regions of the reciprocal space quantum coherence is preserved.
% At the same time, the disappearance of periodic Rabi oscillations and the presence of irregular interference patterns 
Apparently, coherent effects suffer from effective dephasing induced by intraband motion in the presence of multiple bands.
% the charge carriers experience a pseudo-dephasing from the displacement term $\propto \bm F(t)\nabla_{\bm k} \rho_{ij}$ of Eq.~\eqref{eq:model} in presence of other bands.
Similarly to Landau damping \cite{Landau_J_Phys_1946_Landau_Damping}, this phenomenon is not related to electron-electron collisions or interaction with environment.

Our results also highlight the role of symmetries in strong-field phenomena.
For $\varphi_{\text{CE}}=0$, the Hamiltonian is symmetric with respect to time reversal.
For $\varphi_{\text{CE}}=\pm \pi/2$, it is PT-symmetric, that is, invariant under simultaneous parity (P) and time-reversal (T) transformations.
The symmetries of final states match those of the Hamiltonian in two-band simulations (e.g. there is the $\bm{k}\leftrightarrow-\bm{k}$ symmetry for $\varphi_{\text{CE}}=0$).
Adding more bands breaks this apparent relation (see Figs.~\ref{Figure1} and \ref{Figure3}).

\section{Conclusions}

We have shown that resonant interaction of strong ultrashort laser pulses with a solid is determined by interdependent dynamics of interband transitions and intraband electron motion in reciprocal space.
Rapid passage of the efficient transition region (the vicinity of the $\Gamma$ point) brings about a new excitation regime that we call kicked anharmonic Rabi oscillations (KARO).
This regime is not specific to GaAs, and similar dynamics can be driven and controlled using broadband optical waveforms (light transients).
The predicted effects are experimentally observable: the asymmetric momentum distribution can be directly observed using angular resolved photoemission spectroscopy (ARPES) \cite{Damascelli_et_al_RevModPhys.75_2003_ARPES_of_Cuprates, Mathias_RSI_2007, Guedde_Science_2007, Shen_et_al_JOP_2009_ARPES}, and the residual current can be detected through accompanying terahertz radiation \cite{Cote_APL_1999,Spasenovic_PRB_2008}.
Our findings add resonant processes to the toolkit of petahertz solid-state technology where potential applications may range from CEP detection~\cite{PaaschColberg2014} to sub-laser-cycle spectroscopy~\cite{Ivanov_CP_2013} and ultrafast signal processing~\cite{Krausz2014}.

\begin{acknowledgments}
We gratefully acknowledge F.~Krausz for drawing our interest to resonant strong-field interactions and for his helpful comments.
The work of M.\,S.\,W., S.\,Yu.\,K.\ and M.\,C.\ was supported by the DFG Cluster of Excellence: Munich-Centre for Advanced Photonics (MAP).
The work of M.\,I.\,S.\ was supported by the a MURI Grant  \#FA9550-15-1-0037 from the US Airforce Office of Scientific Research.
A support for V.\,S.\,Y.\ came from a MURI Grant \#N00014-13-1-0649 from the US Office of Naval Research.
\end{acknowledgments}

%%%%%%%%%%%%%%%%%%%%%%%%%%%%%%%%%%%%%%%%%%%%%%%%%%%%%%%%%%%%%%%%%%%%%%%%%%%%%%%%%%%%%%%% 

\appendix

\section{Band structure calculations}

\begin{figure}[!htb]
  \centering
  \includegraphics[height=6cm]{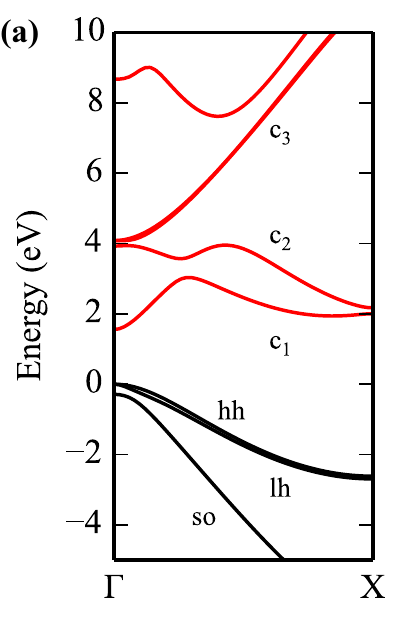}\;
  \includegraphics[height=6cm]{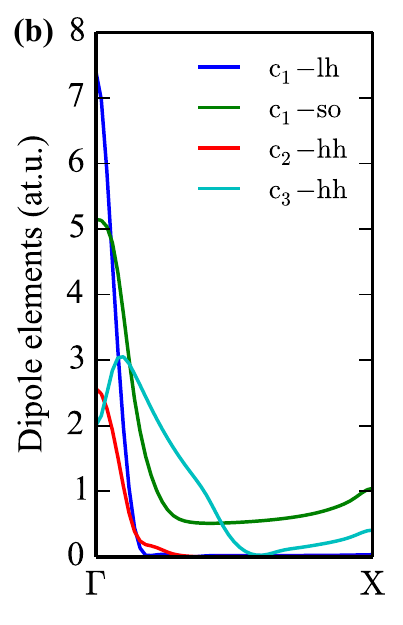}
  \caption{(a) The energies of three highest valence bands (black) and five lowest conduction bands (red) of GaAs along the line between the $\Gamma$ and X points. Each of these bands is doubly degenerate.
    (b) Dipole moments $|\bm{d}_{i j}(\bm{k})|$ for the most important interband transitions.}
  \label{fig:bands}
\end{figure}

Fig.~\ref{fig:bands} shows the band energies and matrix elements of GaAs that we obtained from the density functional theory (DFT) calculations, where we used the TB09 meta-GGA exchange-correlation potential~\cite{Tran2009} with spin-orbit coupling.
The calculated band-gap energy is $E_g=1.55$~eV, which is somewhat larger than the measured band gap at room temperature (1.42~eV) but close to that at low temperatures (1.52~eV)~\cite{Blakemore1982}.

We evaluate the reduced electron mass by fitting the gap between the lowest conduction band ($\text{c}_1$) and the light-hole (lh) band with a second-order polynomial within $|k| \le 0.05 k_{\text{max}}$.
The result is $0.053 m_0$.

The Wien2K code only provides us with momentum matrix elements $\bm{p}_{ij}$, so we evaluated the interband matrix elements of crystal coordinate via $\bm{d}_{i \ne j}(\bm{k}) = \iu e \hbar\bm{p}_{ij}(\bm{k})/\bigl(m_0 [E_{i}(\bm{k})-E_{j}(\bm{k})]\bigr)$, where $m_0$ is the free electron mass.
We found that Berry connections $e^{-1} \bm{d}_{ii}$ had a negligible effect on the residual intraband current density, so we assumed $\bm{d}_{ii} = 0$.

\section{External field}

\begin{figure}[!htb]
  \centering
  \includegraphics[width=0.9\linewidth]{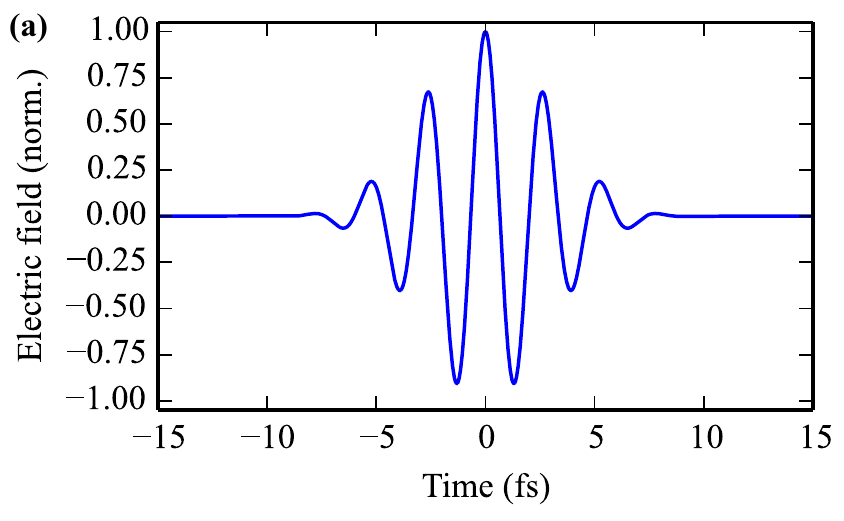}
  \includegraphics[width=0.9\linewidth]{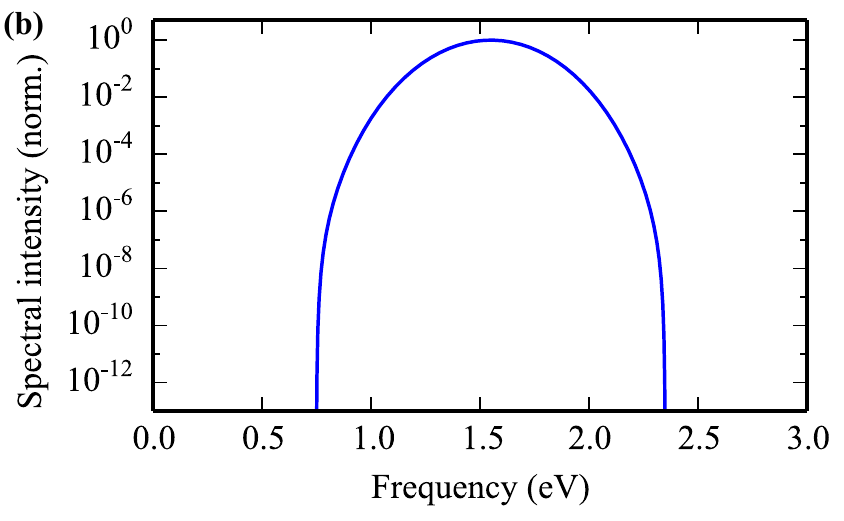}
  \caption{(a) Normalized electric field inside the medium.
    (b) The power spectrum of the laser pulse in the limit of an infinitely broad temporal window.}
  \label{fig:laser}
\end{figure}

Fig.~\ref{fig:laser}(a) shows the field $F(t)$ that we used in our simulations, while Fig.~\ref{fig:laser}(b) shows the power spectrum of the pulse.
We define the field $F(t) = \hat{\bm{e}} \bm{F}(t)$ in the frequency domain taking the spectrum $\tilde{F}(\omega)$ as a Nuttall window centered at $\omega_0 = \hbar^{-1} E_g$ with a compact support in the interval $E_g-\hbar\Delta\omega \le \hbar\omega \le E_g+\hbar\Delta\omega$ with $\hbar\Delta\omega=0.8$~eV.
The central wavelength of the pulse is $\lambda_0 = 2 \pi c / \omega_0 = 800$~nm; its central photon energy is $\hbar\omega_0 = E_g = 1.55$~eV.
In the time domain, the intensity of the pulse has a full width at half maximum (FWHM) of $5$~fs; our simulations begin at $t_0=-36.2$ and end at $t_{\text{max}}=36.2$~fs.
An explicit expression for the positive-frequency part of the pulse spectrum is $\tilde{F}(\omega) = \tilde{F}_0 \exp(\iu \varphi_{\text{CE}}) w\bigl[(\omega-\omega_0+\Delta\omega) / (2\Delta\omega)\bigr]$, where $\varphi_{\text{CE}}$ is the carrier-envelope phase.
The Nuttall window is defined as $w(x) = \sum_{n=0}^3 a_n \cos(2 \pi n x)$ if $0 \le x \le 1$ and $w(x)=0$ outside of this interval. The coefficients are $a_0 = 0.355768$, $a_1 = -0.487396$, $a_2 = 1/2 - a_0$, and $a_3 = -(1/2 + a_1)$.

\section{Temporal evolution}

\begin{figure}[!h]
  \centering
  \includegraphics[width=\columnwidth,height=0.7\textheight,keepaspectratio]{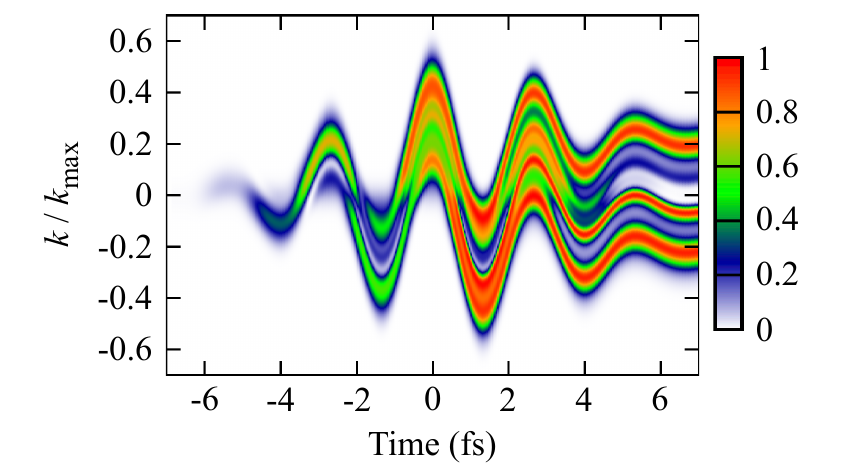}
  \caption{Temporal evolution of the conduction-band population in a two-band simulation for $F_0=0.5$~V/{\AA}, $\varphi_{\text{CE}} = \pi/2$, $T_2 = \infty$.}
  \label{fig:t-k_analysis}
\end{figure}

Fig.~\ref{fig:t-k_analysis} illustrates the temporal evolution of electronic excitations.
Interband transitions mainly take place near the $\Gamma$ point.
Electron-hole wavepackets excited by different laser half-cycles interfere with each other.
This interference determines residual reciprocal-space population distributions.
Asymmetric distributions yield residual ballistic electric current.

\section{Quantum-path interference in reciprocal space}
% \label{sec:QPI}

As explained in the main text, the KARO regime is characterized by ``kick-like'' interband transitions that take place when an electron moving along a reciprocal-space trajectory,
\begin{equation}
\bm{K}(t) = \bm{k} - \frac{e}{\hbar} \int_{t_0}^t \bm{F}(t') \de t',
\end{equation}
traverses a region where the corresponding transition matrix element $\bm{d}_{ij}(\bm{K}(t)\bigr)$ is particularly large.
In this section, we study how such temporally confined transitions interfere with each other.
This task is most easily accomplished if we neglect interband phase relaxation ($T_2 \to \infty$) and solve the time-dependent Schr\"odinger equation in the basis of Houston states.
Let $\ket{\psi_i(\bm{k})}$ be a Bloch state for band $i$ and crystal momentum $\bm{k}$.
Using the ansatz
\begin{equation}
\label{eq:Houston_ansatz}
\ket{\Psi_{\bm{k}}(t)} = \sum_i \alpha_{i,\bm{k}}(t)
\ee^{-\frac{\iu}{\hbar} \int_{t_0}^t \de t'\, E_i\bigl(\bm{K}(t')\bigr)}
\ket{\psi_{i}\bigl(\bm{K}(t)\bigr)},
\end{equation}
where $\alpha_{i,\bm{k}}(t)$ are probability amplitudes, we obtain the following equations~\cite{Bychkov_1970_JETP_31_928,Krieger1986}:
\begin{multline}
\label{eq:Kriegers_equations}
\frac{\de}{\de t} \alpha_{i,\bm{k}}(t) = -\frac{\iu}{\hbar} \sum_j \alpha_{j,\bm{k}}(t)
\bm{F}(t) \cdot \bm{d}_{ij}\bigl(\bm{K}(t)\bigr)\times\\
\exp\left[\frac{\iu}{\hbar} \int_{t_0}^t \de t' \Delta E_{ij}\bigl(\bm{K}(t')\bigr)\right].
\end{multline}
% The choice of initial time $t_0$ is arbitrary---it only adds a constant phase to $\Psi_{\bm{k}}$ and does not affect physical observables.
Ansatz \eqref{eq:Houston_ansatz} accounts for the intraband motion, as well as for the quantum phase accumulated by a wave function in the absence of interband transitions.
Therefore, probability amplitudes $\alpha_{i,\bm{k}}(t)$ only change their values when electrons are excited or de-excited.
Now we introduce the approximation of sudden transitions at a series of moments $\{t_n\}$, which may themselves depend on initial crystal momentum $\bm{k}$:
\begin{equation}
\bm{d}_{ij}(\bm{K}(t)\bigr) \approx \sum_n \delta(t-t_n) \int_{t_n-\Delta t}^{t_n+\Delta t} \de t\, \bm{d}_{ij}(\bm{K}(t)\bigr).
\end{equation}
Here, $\delta(t)$ is the Dirac delta-function, $\left| \bm{d}_{ij}(\bm{K}(t)\bigr) \right|$ has a local maximum at $t=t_n$, and the integration is performed over a short interval of time ($\Delta t \ll t_{n+1} - t_n$) during which transitions are most probable. Let us change the integration variable from time to crystal momentum
assuming that the laser field is linearly polarized along a unit vector $\hat{\bm{e}}$:
\begin{multline}
\int_{t_n-\Delta t}^{t_n+\Delta t} \de t\, \bm{d}_{ij}(\bm{K}(t)  \bigr) =
\int_{\bm{K}(t_n - \Delta t)}^{\bm{K}(t_n + \Delta t)} \frac{\de k}{\left| \bm{K}'(t) \right|}\,
\bm{d}_{ij}(\bm{k}) \approx\\
-\frac{\hbar}{e |F(t_n)|} \int_{-\Delta k}^{\Delta k} \de k\,
\bm{d}_{ij}\left( \bm{K}(t_n) + k \hat{\bm{e}} \right).
\end{multline}
Here, $\Delta k = e \hbar^{-1} |F_0| \sim \left| \bm{K}'(t_n) \right| \Delta t$.
Introducing
\begin{equation}
D_{i j}(\bm{k}, t_n) = \frac{1}{e} \int_{-\Delta k}^{\Delta k} \de k\, \bm{d}_{ij}\left( \bm{K}(t_n) + k \hat{\bm{e}} \right), 
\end{equation}
we re-write Eq.~\eqref{eq:Kriegers_equations} as
\begin{multline}
\label{eq:kicked_Krieger}
\frac{\de}{\de t} \alpha_{i,\bm{k}}(t) \approx \iu \sum_j \alpha_{j,\bm{k}}(t)
\exp\left[\frac{\iu}{\hbar} \int_{t_0}^t \de t' \Delta E_{ij}\bigl(\bm{K}(t')\bigr)\right]\\
\sum_n \delta(t-t_n) \text {sgn}[F(t_n)] D_{i j}(\bm{k}, t_n).
\end{multline}
In this model, the probability amplitudes stay constant between kicks, and they change their values abruptly at each kick.
A general solution of Eq.~\eqref{eq:kicked_Krieger} can be written in the matrix form.
For one particular transition at time $t_n$,
\begin{equation}
\bm{\alpha}_{\bm{k}}(t_n+\Delta t) = \ee^{\iu \hat{M}(\bm{k}, t_n)} \bm{\alpha}_{\bm{k}}(t_n - \Delta t),
\end{equation}
where the elements of matrix $\hat{M}(\bm{k}, t_n)$ are given by
\begin{multline}
M_{i j}(\bm{k}, t_n) = \exp\left[\frac{\iu}{\hbar} \int_{t_0}^{t_n} \de t' \Delta E_{ij}\bigl(\bm{K}(t')\bigr)\right]\\ \times
\text {sgn}[F(t_n)] D_{i j}(\bm{k}, t_n).
\end{multline}
As long as the probability of a particular transition is small, then $\ee^{\iu \hat{M}(\bm{k}, t_n)} \approx 1 + \iu \hat{M}(\bm{k}, t_n)$.
In this approximation,
\begin{multline}
\label{eq:approximate_Delta_alpha}
\Delta \alpha_{i,\bm{k}}(t_n) =
\alpha_{i,\bm{k}}(t_n + \Delta t) - \alpha_{i,\bm{k}}(t_n - \Delta t) \approx\\
\iu \text {sgn}[F(t_n)] \sum_j \Biggl\{ \alpha_{j,\bm{k}}(t_n - \Delta t) D_{i j}(\bm{k}, t_n)\\
\times
\exp\left[\frac{\iu}{\hbar} \int_{t_0}^{t_n} \de t' \Delta E_{ij}\bigl(\bm{K}(t')\bigr)\right]\Biggr\}.
\end{multline}
Whether two particular excitation events at times $t_{n_1}$ and $t_{n_2}$ interfere constructively or destructively depends on the relative phase between $\Delta \alpha_{i,\bm{k}}(t_{n_1})$ and $\Delta \alpha_{i,\bm{k}}(t_{n_2})$, which is given by $\Delta\phi(\bm{k}) = \arg\left[ \Delta \alpha_{i,\bm{k}}^*(t_{n_1}) \Delta \alpha_{i,\bm{k}}(t_{n_2}) \right]$.
If we constraint ourselves to transitions from band $j$ to band $i \ne j$, then Eq.~\eqref{eq:approximate_Delta_alpha} yields
\begin{multline}
\Delta\phi_{i j}(\bm{k}) = \arg\Biggl\{
\text {sgn}[F(t_{n_1}) F(t_{n_2})] \times \\
\alpha_{j,\bm{k}}^*(t_{n_1} - \Delta t) \alpha_{j,\bm{k}}(t_{n_2} - \Delta t)
D_{i j}^*(\bm{k}, t_{n_1}) D_{i j}(\bm{k}, t_{n_2})\\
\times
\exp\left[\frac{\iu}{\hbar} \int_{t_{n_1}}^{t_{n_2}} \de t' \Delta E_{ij}\bigl(\bm{K}(t')\bigr)\right]
\Biggr\}.
\end{multline}
If both transitions take place at the same $\bm{k}$ (e.g.\ $\bm{k} = 0$), then $D_{i j}(\bm{k}, t_{n_1}) = D_{i j}(\bm{k}, t_{n_2})$.
Since we are currently considering small transition probabilities, we can neglect the change of phase of the initial state: $\arg[\alpha_{j,\bm{k}}^*(t_{n_1} - \Delta t) \alpha_{j,\bm{k}}(t_{n_2} - \Delta t)]$.
These considerations lead us to the following result:
\begin{multline}
\Delta\phi_{i j}(\bm{k}) \approx \frac{\pi}{2} \Bigl(1 - \text {sgn}[F(t_{n_1}) F(t_{n_2})] \Bigr) \\
+ \frac{1}{\hbar} \int_{t_{n_1}}^{t_{n_2}} \de t' \Delta E_{ij}\bigl(\bm{K}(t')\bigr).
\end{multline}
The first term on the right-hand side of this equation disappears if $t_{n_1}$ and $t_{n_2}$ are separated by a full laser cycle, and it is equal to $\pi$ if the laser field changes its sign between these two moments of time.

\section{Rabi flopping}

\begin{figure}[!tbp]
  \centering
  \includegraphics[width=\columnwidth,height=0.7\textheight,keepaspectratio]{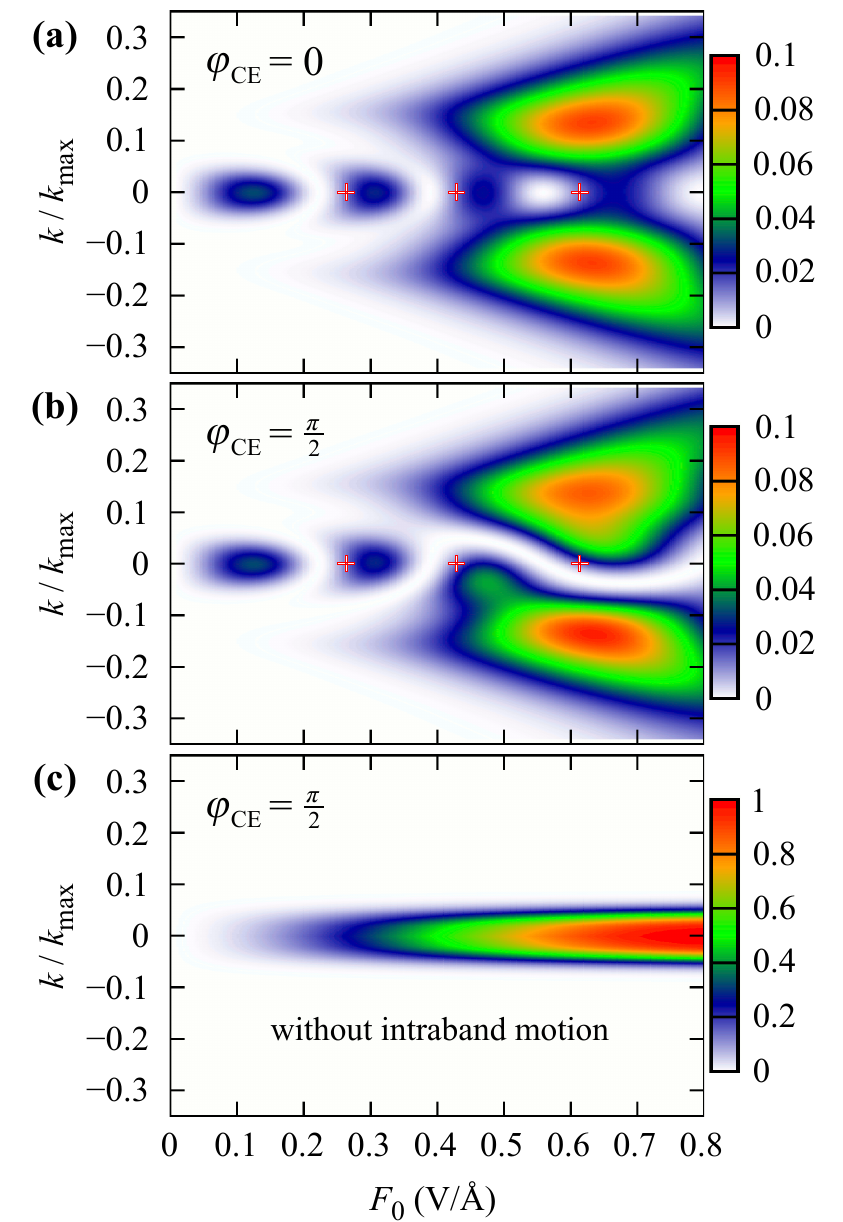}
  \caption{The same as Fig.~\ref{Figure3}, but all the transition matrix elements were reduced by a factor of 10 in order to suppress Rabi oscillations.
  The false colors represent the residual population of conduction-band states in the two-band simulations without dephasing.
  The red crosses indicate the peak field strengths where the pulse area $\mathcal{A}$ [see Eqs.~\eqref{eq:pulse_area}] is an integer multiple of $2\pi$.
  Due to reduced excitation probabilities, panels (a) and (b) use a different color scale.
  }
  \label{fig:population1C1V_no_Rabi}
\end{figure}

As an attempt to disentangle intraband motion from Rabi oscillations, we repeat our simulations with all the transition matrix elements reduced by a factor of 10.
In the absence of detuning, this would decrease the Rabi frequency by the same factor.
Intraband motion causes a large time-dependent detuning enabling interference effects that strongly resemble Rabi flopping, as one can see in Fig.~\ref{fig:population1C1V_no_Rabi}.
There are indeed no signatures of Rabi oscillations if there is no intraband motion [panel (c)]; however, in panels (a) and (b), we see periodic local minima or the conduction-band population at $k=0$.
The field strengths at which these minima occur are slightly larger than those in Fig.~\ref{Figure3} of the main text.
Similarly to Fig.~\ref{Figure3}, strongly asymmetric distributions form at $F_0 \gtrsim 0.4$~V/{\AA}.

To verify that the periodic destructive interference at $k=0$ can be related to Rabi oscillations, we evaluate the pulse area, accounting for the time-dependent detuning induced by the intraband motion:
\begin{subequations}
\label{eq:pulse_area}
\begin{gather}
\mathcal{A} = \int_{-\infty}^{\infty} \de t\,
\sqrt{\left|\Omega_R(t)\right|^2 +
\left[ \Delta E_{ij}\bigl(K(t)\bigr) - \hbar\omega_0 \right]^2},\\
\Omega_R(t) = \frac{d_{i j}\bigl(K(t)\bigr) f(t)}{\hbar},\\
K(t) = \frac{e}{\hbar} A(t).
\end{gather}
\end{subequations}
Here, $f(t)$ is the envelope of the electric field:
\begin{equation}
\label{eq:envelope}
F(t) = 2 \Re\left[f(t) \ee^{-\iu \omega_0 t}\right].
\end{equation}
A long laser pulse with a constant detuning completes an integer number of Rabi cycles if its pulse area is equal to an integer multiple of $2 \pi$.
Even though we consider a few-cycle pulse where detuning $\Delta E\bigl(K(t)\bigr) - \hbar\omega_0$ significantly changes within a fraction of a cycle, it is still instructive to evaluate the pulse amplitudes that satisfy $\mathcal{A} = 2 \pi n$ ($n \in \mathbb{N}$).
These field strengths are indicated in Fig.~\ref{fig:population1C1V_no_Rabi} by red crosses, the corresponding values being 0.26, 0.43, and 0.61~V/{\AA}.
The same analysis applied to the original matrix elements yields peak field strengths of 0.16, 0.33, and 0.48~V/{\AA}, which agree well with the results shown in Fig.~\ref{Figure3} of the main text.
Therefore, in these two examples (Figs.~\ref{fig:population1C1V_no_Rabi} and 3), kicked Rabi oscillations may also be interpreted as strongly detuned Rabi flopping.
However, we do not know if Eq.~(\ref{eq:pulse_area}a) is generally applicable in the KARO regime.
Also, unlike the analysis presented in the previous section, Eqs.~\eqref{eq:pulse_area} is inapplicable to very broadband pulses where the central frequency is not well-defined and, therefore, Eq.~\eqref{eq:envelope} does not provide an unambiguous definition of pulse envelope.

\begin{figure}[!htb]
  \centering
  \includegraphics[width=\columnwidth,height=0.7\textheight,keepaspectratio]{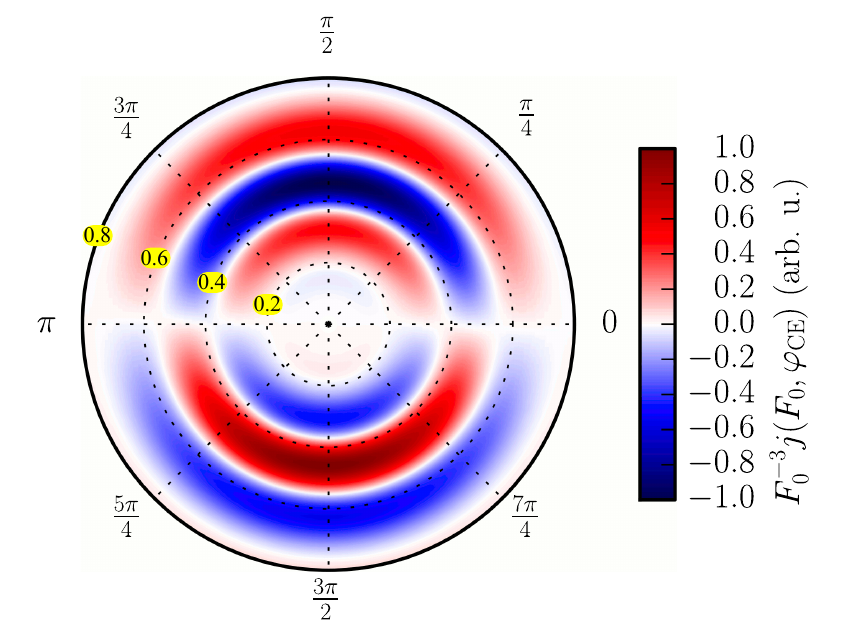}
  \caption{Residual electric current evaluated with the transition matrix elements being reduced by a factor of 10. Here, we used two bands and $T_2=\infty$.
  }
  \label{fig:current3C3V_no_Rabi}
\end{figure}

Figure~\ref{fig:current3C3V_no_Rabi} shows that completing a Rabi cycle causes current reversal even when the transition matrix elements are reduced.

\section{Transferred charge}

\begin{figure}[!htbp]
  \includegraphics[width=\columnwidth,height=0.7\textheight,keepaspectratio]{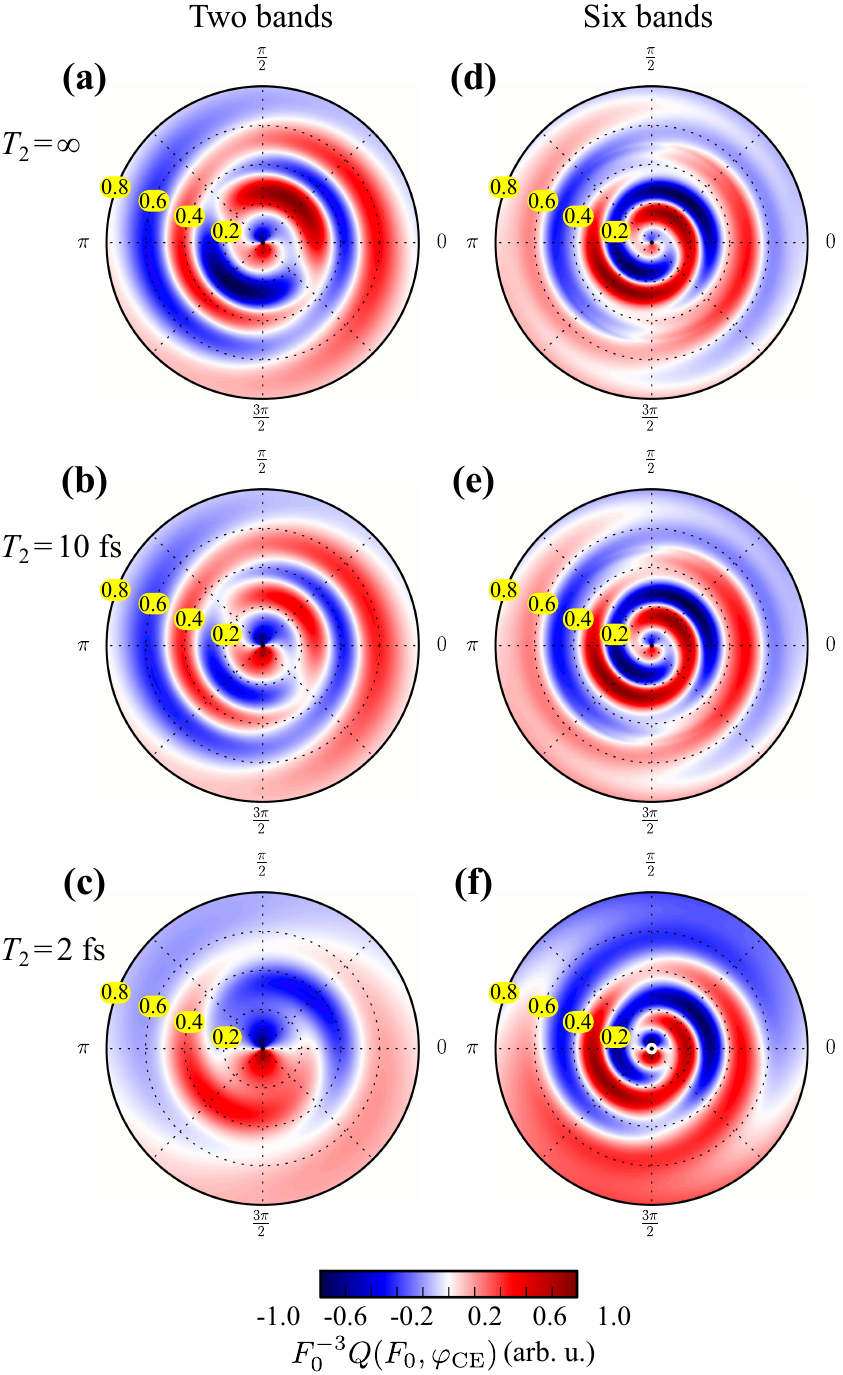}
  \caption{The transferred charge density $Q(F_0, \varphi_{\text{CE}})$.
    In these diagrams, the distance to the origin corresponds to pulse amplitude $F_0$, which varies from zero to 0.8~V/{\AA}, while the angle to the horizontal axis encodes the carrier-envelope phase $\varphi_{\text{CE}}$.
    The false colors represent $F_0^{-3} Q(F_0, \varphi_{\text{CE}})$ individually normalized for each diagram.
%    The solid green curves represent $\varphi_{\text{CE}}$ that maximizes the transferred charge.
    The left three panels show two-band results (one valence and one conduction band), while panels (d)-(f) show results obtained with three valence and three conduction bands.
    Each horizontal pair of plots corresponds to a certain value of the dephasing time $T_2$ as indicated by the labels.
  }
  \label{fig:Q}
\end{figure}

In this paper, we focused on the residual current density because this quantity is directly related to asymmetry in the residual band population, and also because it was primarily used in the previous work on $\omega+2\omega$ interference.
However, for an experimental verification of these results, the transferred charge may be a more relevant observable~\cite{Schiffrin2013,Kruchinin_PRB_2013,PaaschColberg2014}.
Fig.~\ref{fig:Q} is similar to Fig.~\ref{Figure1} of the main text, but, instead of the residual current, we visualize the transferred charge density
\begin{equation}
  \label{eq:Q}
  Q\left(F_0, \varphi_{\text{CE}}\right) = \int_{-\infty}^{t_{\text{cut}}}
  \de t\, j\left(F_0, \varphi_{\text{CE}}, t\right),
\end{equation}
where we set the upper integration limit to $t_{\text{cut}} = 8$~fs  and evaluate the time-dependent intraband current density according to
\begin{equation}
  j\left(F_0, \varphi_{\text{CE}}, t\right) =
  -\frac{2 e}{(2 \pi)^3}\sum_i \int_{\text{BZ}} \de^3 k\, n_i(\bm{k}, t)
  \hat{\bm{e}} \bm{v}_i(\bm{k}).
\end{equation}

%%%%%%%%%%%%%%%%%%%%%%%%%%%%%%%%%%%%%%%%%%%%%%%%%%%%%%%%%%%%%%%%%%%%%%%%%%%%%%%%%%%%%%%% 

\bibliographystyle{apsrev4-1}
\bibliography{Wismer_et_al}

\end{document}